\documentstyle[aabib,psfig]{l-aa}

\thesaurus{05;
           03.13.4; 
           05.03.1;
           08.05.3;
           10.07.2;
           10.15.1}
\newcommand{\msun}{\mbox{${\rm M}_\odot$}}

\newcommand{\Kms}{\mbox{${\rm km~s}^{-1}$}}

\newcommand{\kira}{\mbox{${\sf kira}$}}
\newcommand{\SeBa}{\mbox{${\sf SeBa}$}}
\def\unit#1{{\mbox{[{\rm #1}]}}}
\def\aplt{\ {\raise-.5ex\hbox{$\buildrel<\over\sim$}}\ }

\def\cut#1{{}}

\begin{document}

\title{On the dissolution of evolving star clusters}

\author{Simon F.\ Portegies Zwart\thanks{Japan Society for the
Promotion of Science Fellow}\inst{1, 2}
\and    Piet Hut\inst{3}	
\and 	Junichiro Makino\inst{2}
\and    Stephen L.\ W.\ McMillan\inst{4}}

\offprints{Simon Portegies Zwart: spz@grape.c.u-tokyo.ac.jp}
\institute{$^1$ Astronomical Institute {\em Anton Pannekoek}, 
                Kruislaan 403, 1098 SJ, Amsterdam, NL \\
	   $^2$ Department of Information Science and Graphics, 
		College of Arts and Science, 
		University of Tokyo, 3-8-1 Komaba,
		Meguro-ku, Tokyo 153, Japan \\
           $^3$ Institute for Advanced Study,
                Princeton, NJ 08540, USA \\
           $^4$ Department of Physics and Atmospheric Science, 
		Drexel University, 
                Philadelphia, PA 19104, USA}

\date{received; accepted:}
\maketitle
\markboth{Portegies Zwart et al.\, :Star Cluster Dissolution}{}

\begin{abstract}

Using direct N-body simulations which include both the evolution of
single stars and the tidal field of the parent galaxy, we study the
dynamical evolution of globular clusters and rich open clusters. We
compare our results with other N-body simulations and Fokker-Planck
calculations.  Our simulations, performed on the GRAPE-4, employ up to
32,768 stars.  The results are not in agreement with Fokker-Planck
models, in the sense that the lifetimes of stellar systems derived
using the latter are an order of magnitude smaller than those obtained
in our simulations.  For our standard run, Fokker-Plank calculations
obtained a lifetime of 0.28 Gyr, while our equivalent $N$-body
calculations find $\sim4$ Gyr.  The principal reason for the
discrepancy is that a basic assumption of the Fokker-Plank approach is
not valid for typical cluster parameters.  The stellar evolution
timescale is comparable to the dynamical timescale, and therefore the
assumption of dynamical equilibrium leads to an overestimate of the
dynamical effects of mass loss.  
Our results suggest that the region
in parameter space for which Fokker-Planck studies of globular cluster
evolution, including the effects of both stellar evolution and the
galactic tidal field, are valid is limited.
The discrepancy is largest for clusters with short lifetimes.

\end{abstract}
	  
\section{Introduction}\label{sec_introduction}

Theoretical models for the evolution of star clusters are generally
too idealized for comparison with observations.  However,
detailed model calculations with direct $N$-body methods are not
feasible for real globular clusters, even with fast special-purpose
computers such as GRAPE-4 (Makino et al.\
1997)\nocite{1997ApJ...480..432M} or advanced parallel computers
(Spurzem \& Aarseth 1996).\nocite{1996MNRAS.282...19S}

If we could scale the results of $N$-body simulations with relatively
small numbers of particles (such as $\sim30,000$) to real globular
clusters, then it would become feasible to perform computations with
relatively small numbers of particles and still derive useful
qualitative conclusions about larger, more massive systems.  However,
to determine the proper scaling is difficult because the ratio between
two fundamental time scales, the relaxation times and the dynamical
time, is proportional to $N$.  In typical globular clusters, this
ratio exceeds $10^3$ and the two time scales are well separated.  In
$N$-body simulations, the ratio is generally much smaller.

The inclusion of realistic effects such as mass loss due to stellar
evolution and the effect of galactic tidal fields (with the galaxy
approximated as a point mass, but also with the inclusion of disc
shocking) further complicate the scaling problem (see, e.g., Heggie
1996).\nocite{heg96} A proper treatment of stellar evolution is
particularly problematic, since its characteristic timescale changes
as stars evolve.

Chernoff \& Weinberg (1990, CW90)\nocite{cw90} performed an extensive
study of the survival of star clusters using Fokker-Planck
calculations which included 2-body relaxation and some rudimentary
form of mass loss from the evolving stellar population.  In their
simulations the number of particles is not specified.  Their models
are defined by the initial half-mass relaxation time and by the
initial mass function of the cluster.  Since their models do not
specify the number of stars per cluster, each of their model
calculations corresponds to a one-dimensional series of models, when
plotted in a plane of observational values, such as total mass versus
distance to the galactic center (Fig.\ 1).  All points of the solid
line in that figure correspond to a single calculations by CW90, since 
they have an identical relaxation time.  As we will see later, it is
useful to consider other series of models, for which the crossing time 
is held constant while varying the mass.  An example of such a series
is indicated by the dashed line in Fig.\ 1.  The shapes of these lines
are derived under the assumption of a flat rotation curve for the
parent galaxy.

The main conclusion of CW90 was that the majority of the simulated
star clusters dissolve in the tidal field of the galaxy within a few
hundred million years.
Fukushige \& Heggie (1995, FH95)\nocite{fh95} studied the evolution
of globular clusters using direct $N$-body simulation, using the same
stellar evolution model as used by CW90. They used a maximum of 16k
particles and a scaling in which the dynamical timescale of the
simulated cluster was the same as that of a typical globular cluster,
corresponding to one of vertical lines in Fig. 1.

FH95 found lifetimes much longer than those in CW90's Fokker-Planck
calculations, for the majority of the models used in CW90.  However,
the reason for the discrepancy is rather unclear, because the
calculations of FH95 and those of CW90 differ in several important
respects.  The relaxation times differ because FH95 held the cluster
crossing time fixed in scaling from the model to the real system.
However, the crossing times themselves are also different, since the
crossing time is by definition zero in a Fokker-Planck calculation.
Finally, the implementation of the galactic tidal field is also quite
different.  CW90 used a simple boundary condition in energy space
(spherically symmetric in physical space), in which stars were removed
once they acquired positive energy, but the underlying equations of
motion included no tidal term.  FH95 adopted a much more physically
correct treatment, including tidal acceleration terms in the stellar
equations of motion and a proper treatment of centrifugal and coriolis
forces in the cluster's rotating frame of reference (see FH95).

\begin{figure}
\centerline{
\psfig{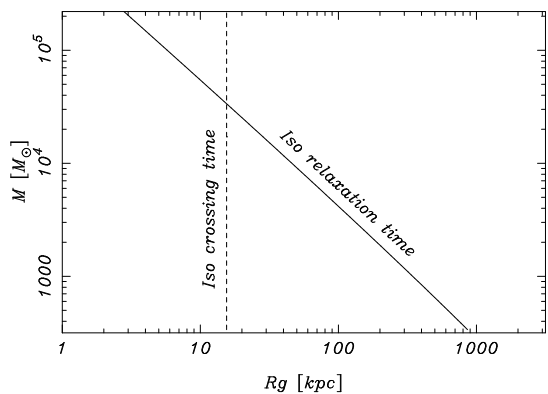}}
\caption
{Cluster mass versus the distance to the galactic center.  The solid
line indicates the model parameters for which the relaxation time is
constant (iso relaxation time); the dashed line indicates the initial
conditions for which the crossing time of the star cluster is constant
(iso crossing time) }
\label{isomodels}\end{figure}

In order to study the behavior of star clusters with limited numbers
of stars, and to compare with the results of the Fokker-Planck
simulations of CW90, we selected one of their models and perform a
series of collisional N-body simulations in which the evolution of the
individual stars is taken into account.  According to CW90 the results
should not depend on the number of stars in the simulation as long as
the relaxation time is taken to be the same for all models. It is,
among others, this statement which we intend to study.  We find that
for this set of initial conditions Fokker-Planck models do not provide
a qualitatively correct picture of the evolution of star clusters.
The effects of the finite dynamical time scale are large, even for
models whose lifetime is several hundred times longer than the
dynamical time.

The main purpose of this paper is to study the survival probabilities
of star clusters containing up to a few tens of thousands of single
stars, in order to gain a deeper understanding of the influence of the
galactic tidal field and the fundamental scaling of small $N$ clusters
to larger systems.  Only single stars are followed; primordial
binaries are not included.  The computation of gravitational forces is
performed using the GRAPE-4 (GRAvity PipE, see \cite{emf+93}, a
special-purpose computer for the integration of large collisional
$N$-body systems).  Hardware limitations (speed as well as storage)
restrict our studies to $\aplt 32$k particles.

The dynamical model, stellar evolution, initial conditions and scaling
are discussed in Sect. 2.  Section 3 reviews the software environment
and the GRAPE-4 hardware, and discusses the numerical methods used.
The results are presented in Sect. 4 and discussed in Sect. 5; Sect. 6
sums up.

\section{The model} 

The simulations presented in this paper were performed by a computer
code that consists of two independent parts.  One part, the $N$-body
model, integrates the equations of motion of all individual stars
while the other part computes the evolution of each single star (see
Sect.~\ref{sec:the_comp_env} for a description of the numerical
implementation).

\subsection{Stellar dynamics}
The equations of motion of the stars in the stellar system are
computed using Newtonian gravity. The numerical integration is
performed using a fourth-order individual--time-step Hermite scheme
(Makino \& Aarseth 1992; see Sect. 3 below).\nocite{ma92}

The radius of the star cluster is limited by the galactic tidal field.
We assume for simplicity that the cluster describes a circular orbit
around the galactic center.  At fixed time intervals, stars that are
outside the tidal radius are simply removed.  We chose this simple
cutoff in order to facilitate direct comparison with the Fokker-Planck
results.  Shocks due to the passage of the cluster through the
galactic plane, twice per orbit, and encounters with giant molecular
clouds are neglected.

At regular time intervals, the stellar evolution model updates all
stars to the current age of the stellar system and the $N$-body model
is notified of the changes in mass and radius of the stars.  Whenever
a star loses more than 1\% of its mass, the dynamical part of the
simulation is reinitialized, to take into account the loss of mass and
energy from the system.  A treatment of collisions and mergers
involving two or more stars is also included.

\subsection{Stellar evolution}

\begin{figure}
\centerline{
\psfig{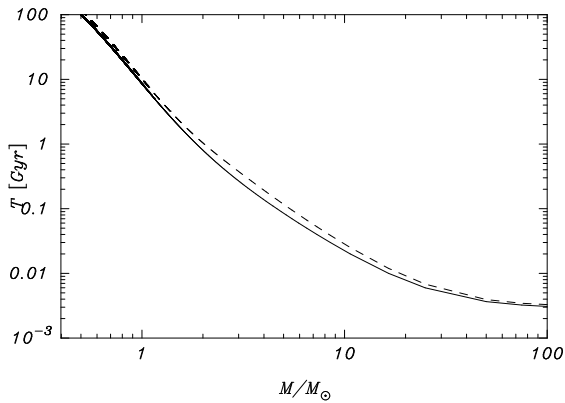}}
\caption{Main-sequence lifetime (solid line) and terminal age (dashed
line) in billions of years for stars as a 
function of the zero age mass of the stellar
evolution tracks of Eggleton et al.\ (1989).}
\label{fig:ppe_mto}\end{figure}

The evolution of single stars in our model is performed using the
stellar evolution model presented by Portegies Zwart \& Verbunt 1996,
and later dubbed \SeBa\ (see Portegies Zwart at al.~1997a and 1997b for a
dynamical
implementation).\nocite{pzhv97}\nocite{pzhmv97}\nocite{pzv96} These
models are based on fitting formulae to stellar evolution tracks of
population~I stars given by Eggleton et al. (1989)\nocite{eft89}
(evolutionary tracks for population~II stars are not yet available in
this convenient form).  These equations give a star's luminosity and
temperature as functions of age and initial mass.  Other stellar
parameters---radius, mass loss and the mass of the core---are then
from these derived.  Figure~\ref{fig:ppe_mto} gives the total and
main-sequence lifetime of the stars in the model as a function of
their mass at zero age.

We point out here a few important details of the evolutionary model: A
star with a mass larger than 8~\msun\ becomes a neutron star in a
supernova, less massive stars become white dwarfs.  Neutron stars have
a mass of 1.34~\msun; the mass of a white dwarf is given by the core
of the star as it leaves the end of the supergiant branch.

Mass lost by stellar evolution, either in the form of a stellar wind
or after a supernova, escapes from the stellar system, as the velocity
of the stellar wind or supernova shell exceeds the escape velocity of
the star cluster.  The escaping mass is presumed to carry away the
same specific energy and orbital angular momentum as the mass-losing
star has.

\subsection{Initial conditions} \label{sec:initial_conditions}

For the initial model, we followed the specifications of CW90.  Each
simulation is started at $t=0$ by giving all stars a mass $m$ drawn
 from a single-component power-law mass function of the form $dN(m) =
m^{-\alpha}dm$, between lower and upper mass limits $m_-$ and $m_+$,
respectively.  From the mean mass $\langle m \rangle$, given by the
mass function, and the total number of stars in the computation $N$,
the initial mass of the star cluster is computed: $M = N \langle m
\rangle$.

The initial density profile and the velocity distribution of the
stellar system are taken from King models (King 1966)\nocite{kin66}.
In a King model, the dimensionless depth of the central potential
$W_\circ$ determines the structure of the cluster and thus the ratio
of the virial radius $r_{\rm vir}$ to the tidal radius $r_t$ (see
e.g.\ Binney \& Tremaine 1987)\nocite{bt87}.  The velocity
distribution is given by a lowered Maxwellian with the same formal
velocity dispersion for all stars, independent of mass.  The initial
positions of the stars are chosen independent of their mass.

The tidal radius is computed assuming that the star cluster initially
fills its Roche lobe in the tidal field of the galaxy.  In other
words, we take the tidal radius of the initial King model as the
physical tidal cutoff radius (Takahashi et al.\ 1997).\nocite{tli97}
The mass of the galaxy $M_G$ and distance $R_G$ from the
cluster to the center of the galaxy are related by assuming that the
star cluster has a circular orbit around the galactic center with a
velocity of $v_G = 220\Kms$:
\begin{equation}
	M_G(R_G) = {v^2_G R_G \over G}.
\label{eq:velocity}\end{equation}

We approximate the tidal radius $r_t$ as the Jacobi radius for the
star cluster.  The distance from its center of mass to the first
Lagrangian point is:
\begin{equation}
r_t(R_G) = \left(M \over 3M_G\right)^{1/3}R_G.
\label{eq:tidal_radius}\end{equation}

After each diagnostic output, the new tidal boundary of the star
cluster is determined, stars outside the tidal boundary are removed
from the stellar system.  As in CW90, no tidal force is applied to the
individual members of the star cluster.  The mass of the star cluster
declines during its evolution due to mass loss from stellar evolution
of the stars and from escaping stars; the tidal radius decreases
accordingly.

All our model clusters start in virial equilibrium.  The computations
are terminated if ten particles remain in the stellar system. The
lifetime refers to the age at which the number of stars in the cluster
is reduced to this limit.

\subsection{System of units}
The system of units used in the $N$-body model are given by
$M = G = -4E = 1$, where E is the initial internal energy of the
stellar system (Heggie \& Mathieu 1986). \nocite{hm86} The
transformation from these scaled $N$-body units to physical units is
realized with a set of transformations for mass, length and time.

The total mass of the stellar system determines the mass scaling.
Since the star cluster starts filling its Roche lobe, the size scaling
is determined by the tidal radius of the initial King model, which is
in turn set by the tidal radius of the star cluster in the galactic
tidal field.  
In this paper, we have adopted $\tau$ as the definition of the
unit of time which is of order unity in our natural system of
units ($2\sqrt{2}$ of the real crossing time).
This corresponds to
\begin{equation}
\tau = 15 \left( {[\msun] \over M} \right)^{1/2}
             \left( {r_{\rm vir}\over [{\rm pc}]}\right)^{3/2} \, \unit{Myr}.
\label{eq:crossing_time}\end{equation}
Here $r_{\rm vir}$ is the virial
radius of the stellar system.  Note that $r_{\rm vir} = 1$ is our standard
units, whereas the relation between $r_{\rm vir}$
and $r_t$ is determined by the King parameter $W_\circ$.

The relaxation time of the cluster is defined as
\begin{equation}
t_{rlx} = 2.0 \left( {M \over [\msun]}              \right)^{1/2}
	      \left( {r_{\rm vir} \over [{\rm pc}]} \right)^{3/2} 
	\;\; 	{[\msun] \over M} \, {N \over \log N} 
	\;\;\;\unit{Myr}.
\label{eq:relaxation_time}\end{equation}

\section{Numerical method and its validation} 
\label{sec:the_comp_env}
\subsection{Software} 

For all runs, we used the integration program \kira\ in the Starlab
software toolset (McMillan \& Hut 1996).\nocite{mh96}
Starlab is a collection of utility programs developed for numerical
simulation of star clusters and the data analysis of the results of the
simulation. Utility programs in Starlab are divided into four classes:
(a) programs which create $N$-body snapshots, (b) programs which apply
some transformation to $N$-body snapshots, (c) programs which perform
time integration, and (d) programs which analyze $N$-body
snapshots. The I/O interface of all of these programs is unified in
such a way that the output of one program can always be used as the
input for another program through the UNIX pipe mechanism.

The integration program \kira\ performs the online analysis needed in
this work.  The time integration scheme used in \kira\ is the
individual timestep with a 4th-order Hermite scheme (Makino \& Aarseth
1992), which is essentially the same as the one used in Aarseth's
NBODY4 and NBODY6. What sets \kira\ apart from these other programs is 
the
way it handles close encounters and compact subsystems. In NBODYx, the
subsystems are handled by various regularization techniques according
to the number of particles and their dynamical states.  In contrast,
\kira\ models any compact subsystem as a local hierarchical binary
tree.  This procedure provides a great algorithmic simplification,
without significant loss in either accuracy or speed.

Another difference between \kira\ and the older NBODYx codes concerns
the data structures used. \kira\ is written in an object-oriented
style in which each star is an object with properties which represent
the dynamics (mass in dynamical units, position, velocity etc.) and
properties which represent the stellar evolution (mass in solar
masses, age, effective temperature, luminosity, state of the star,
chemical composition and so on). The part of the code which handles
the dynamics and the part that handles the stellar evolution are
completely independent of each other.  The communication between the
two parts can take place only locally, for one individual star at a time,
and only at one well-defined point in the code, where the dynamical
and evolutionary information of that star are allowed to exchange
information. This design makes it possible to combine complex $N$-body
integration programs and complex stellar evolution packages without
major problems in either compatibility or maintenance.

Currently, the stellar evolution is handled in the following way.  At
a pre-specified time interval, the diagnostic time step (see Fig. 3),
the stellar evolution package \SeBa\ is called for all stars in the
system and the mass and state of the stars are updated. If the change
of the mass of any star exceeds a prescribed limit, the mass of all
stars used in the $N$-body integrator is updated, the energy change due to
that change is calculated, and the integrator is re-initialized.

Since we invoke stellar evolution and tidal truncation only at fixed
time intervals, the results of the calculation might conceivably
depend on the specific time interval used. We discuss this possibility
below, in Sect.\ 3.3.

\subsection{The hardware} 
The time integration in \kira\  for this study is  performed using the
special purpose computer GRAPE-4 (Makino et al. 1997). 	GRAPE-4 is an
attached processor which calculates the gravitational interactions
between stars. The actual time integration is done on the host
computer, which runs the UNIX operating system. Starlab programs
and packages all reside on the host computer.

The peak speed of the fully configured GRAPE-4 system is 1.08 Tflops. 
For this study, we mostly use the smallest configuration with a peak 
speed of 30 Gflops. 

\subsection{Test of the method}

In this section, we describe the result of the test runs in which we
varied the time intervals to invoke the stellar evolution package and
tidal removal of stars. As we mentioned above, this choice could have
a large effect, in particular at the early stage where the
high-mass stars, which evolve very quickly, dominate the evolution of
the stellar system. 

\begin{figure}
\centerline{
\psfig{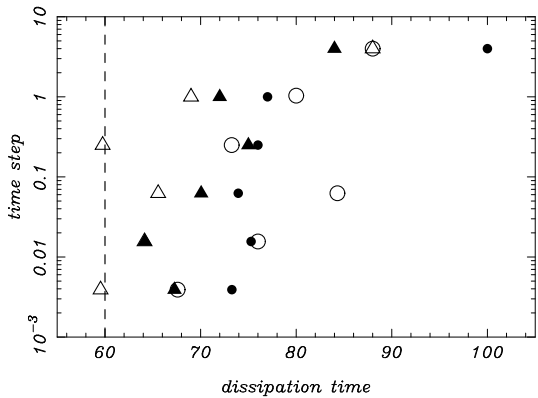}}
\caption{Life time of the star clusters from the test runs with 
1k (triangles) and 4k (circles) stars (see Table~\ref{tab:testruns})
as a function of the diagnostic integration
time step (both in units of the initial crossing time of the stellar
system). The filled symbols refer to the computations with the same
initialization, open symbols are from different random initializations.
The vertical dashed line indicates the result from 
FH95 for a run with 8k stars.}
\label{test_runs}\end{figure}

We performed the simulation from the same initial conditions for 1024-
and 4096-body systems, changing the time interval to invoke the mass
loss due to stellar evolution from 1/256 to 4. The time interval to
remove the stars out of the tidal radius is taken to be the same as
the time interval for stellar mass loss.  All these runs
resulted in the evaporation of the cluster in less than 100~Myrs.
Table~\ref{tab:testruns} and Fig.~\ref{test_runs} give overviews of
the results.

If the stellar evolution time interval is taken to be larger than the
crossing time, then the lifetime of the stellar system is extended
considerably. On the other hand, as long as the evolution time step is
taken to be a small fraction of the crossing time, its precise choice
of value has a relatively small effect on the lifetime of the cluster
and run-to-run variations due to small number statistics are larger
than any systematic effects caused by the choice of time interval.

In these models, the crossing time corresponds to 1.7~Myr. If the
diagnostic time interval is chosen to be longer than the crossing
time, it would exceed the lifetime of the most massive stars.  This in 
turn would delay the effects of stellar evolution to be felt
dynamically, which would artificially extend the lifetime of the
cluster.  This effect is illustrated in Fig.~\ref{test_runs}.

\begin{table}
\caption[]{Series of test runs with 1k and 4k particles of star
clusters with the following initial conditions $W_\circ=3$, $\alpha =
1.5$ between 0.4 and 14\msun and $r_{\rm vir} = 3.2$~pc and $r_{\rm
vir} = 5.1$~pc for the runs with 1k and 4k stars, respectively (where
we keep the crossing time fixed, at $\sim 1.7$~Myr). The numbers in
the table are in units of the $N$-body time.  The first column gives
the name of each test model.  The second column gives the diagnostic
time step for synchronization of the dynamical integrator and the
stellar evolution part of the code.  The next two columns give the
time (in units of a $N$-body time) in which the 1k models dissolves,
followed by two columns that display the time in which the 4k models
dissolve.  The columns listed as ``1k'' and ``4k'' refer to
calculations in which the exact same initial conditions were used
throughout a single column.  The columns listed with an extra ``(r)''
refer to calculations for which each run was started from a new,
random realization. }
\begin{flushleft}
\begin{tabular}{l|llrrr}\hline 
model & $dt$      & 1k    & 1k(r)        & 4k      & 4k(r) \\ \hline
T1    & 1/256	  & 67    & 60           & 73      & 68 \\
T2    & 1/64	  & 64    & 64           & 75      & 76 \\
T3    & 1/16	  & 70    & 66           & 74      & 84 \\
T4    & 1/4	  & 75    & 60           & 76      & 73 \\
T5    & 1	  & 72    & 69 		 & 77      & 80 \\
T6    & 4         & 84    & 88           & 100     & 88 \\ \hline
\end{tabular}
\end{flushleft}
\label{tab:testruns} \end{table}

For the following computations we adopt a
diagnostic timestep for
synchronization of the dynamical integrator and the
stellar evolution part of the code of $dt = 1/64$. This suffices in
terms of accuracy without too much performance lost.

\begin{table}
\caption[]{The computations in which we attempt  to reproduce the
results from FH95 
are based on a star cluster of family 1 (see CW90),
with a total mass of $1.49 \cdot 10^5$\msun.  The mass function is given by
$\alpha = 2.5$ between 0.4\msun and 14\msun and $W_\circ=3$.
The time unit $\tau$ is about 1.9~Myr and the distance to
the galactic center is about 4~kpc.
The first column gives the model name, followed by the number of
particles in the simulation, the tidal radius, the virial radius,
and  the time for dissolving the system in the
tidal field of the galaxy in crossing times and in billion years.
}
\begin{flushleft}
\begin{tabular}{lr|rrrl} \hline
model& $N$  & $r_t$  & $r_{\rm vir}$ & \multicolumn{2}{c}{$t_{diss}$} \\
     &      & \unit{pc}&\unit{pc}& [$\tau$] & \unit{Gyr}\\ \hline
FH1  & 1k   & 8.0    &  2.58     & 220    & 0.42 \\
FH2  & 2k   & 10.1   &  3.31     & 420    & 0.83 \\
FH4  & 4k   & 12.7   &  4.02     & 690    & 1.3  \\
FH8  & 8k   & 16.0   &  5.12     & 1000   & 1.9  \\
FH16 & 16k  & 20.1   &  6.36     & 1100   & 2.0  \\ \hline
\end{tabular}
\end{flushleft}
\label{tab:fh95f7} \end{table}

We performed a second set of test runs using the same initial
conditions as one of the runs in FH95 (the run shown in 
their Fig.\, 7).  The results of our test runs is given in 
Table~\ref{tab:fh95f7}. There are two reasons to perform this test,
namely to investigate the effect of the rather large softening used by 
FH95, and to evaluate to what extent different treatments of a tidal
limit influence the value of the final dissipation time.

\begin{figure}
\centerline{
\psfig{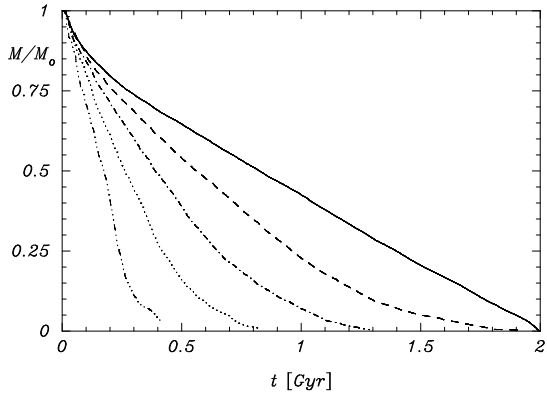}}
\caption{The time evolution of the total mass of the models FH1 (dash-3dot)
line), FH2 (dotted), FH4 (dash dot), FH8 (dashed line) and
FH16 (solid line).}
\label{FH95_mn_vs_time}
\end{figure}

Figure~\ref{FH95_mn_vs_time} demonstrates that the evolution of the
total mass is qualitatively different for runs with less than 16k
particles and the run with 16k particle.  In FH95, this transition
took place around $N=4096$. This result is quite natural since the
relaxation effect is 2-3 times smaller in the FH95 runs, because of
the large softening.

\begin{figure}
\centerline{
\psfig{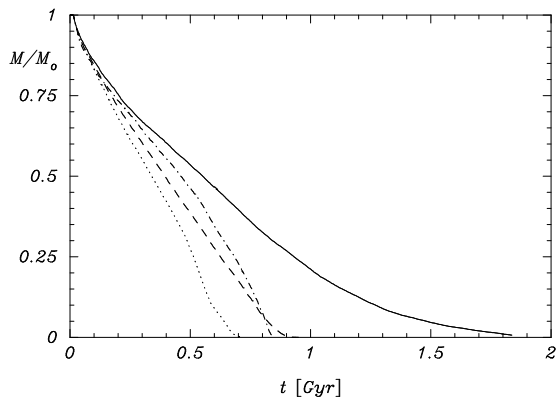}}
\caption{The time evolution of the total mass for models similar to
model FH8 but with various choices for the tidal field and softening.
The solid line gives the results of model FH8 with a tidal cut off and
without any softening (similar to the dashed line in
Fig.\,\ref{FH95_mn_vs_time}). The dashed line gives the result for the
more elaborate implementation of the tidal field (no softening). The
dash-dotted and the dotted line give the results for the softened
models with a tidal cut-off and with the tidal field, respectively.
For each line the average of two runs with identical initial
conditions are selected (with a typical run-to-run variation of
lifetimes of only a few percent).  }
\label{FH95_8k_tide}
\end{figure}

Our cluster dissolution times turn out to be roughly a factor of two
longer than those reported by of FH95. There are two reasons for this
discrepancy, as can be read from Fig.~\ref{FH95_8k_tide}.  One
reason is the fact that we do not use any softening in our
production runs.  Adding a softening, comparable to that used by FH95, 
cuts our dissolution time roughly in half.  Another reason stems from
the difference in treatment of tidal limitations of the model cluster.
Our choice of a simple tidal cutoff, rather than the
more elaborate and physically correct tidal field employed by FH95,
effectively weakens the tidal effect in our calculations,
increasing the cluster lifetime.  

Figure~\ref{FH95_8k_tide} compares several realizations of run FH8
(Table~\ref{tab:fh95f7}), some using our simple spherical cutoff,
others with a tidal field equivalent to that employed by FH95.  All
runs were performed using \kira.  When both changes are made, adding a
large softening as well as a more accurate treatment of the tidal
field (dotted line), the dissolution time diminishes even further, to
less than 40\% of that of our normal run (solid line).

FH95's treatment of the tidal field is physically correct, but it
complicates comparison of the $N$-body results with the CW90's
Fokker-Plank results.  For that reason, in the remainder of this
comparative paper, we continue to use a simple tidal cutoff.  However,
the reader should note that the dissolution times reported below are
probably too long by a factor of two.

\subsection{Initial IC and IR models}

We consider the King model with $W_\circ=3$ with an initial half-mass
relaxation time of 2.87 billion years (family 1 of CW90) as our
standard model.  (Note here that the relaxation time used by CW90 is
defined at the tidal radius of the star cluster instead of at the
virial radius, see Eq.\ref{eq:relaxation_time}.)  The
slope of the initial mass function is fixed to $\alpha = 2.5$. The
cutoff values of the mass distribution at high- and low-mass end are
taken as 14 \msun\ and 0.4 \msun, respectively. These values are
chosen so that they are the same as the parameters used in CW90.
(Note that CW90 quote both 14~\msun\ and 15~\msun\ as upper limits for
their initial mass function.)

The number of particles we used for the standard model is 32768. 
Table~\ref{tab:n32} summarizes the characteristics of the standard
model.

Starting from the standard model, we generated two series of initial
models.  We designed the first series of runs (referred to below as
iso-relaxation, or IR, models) so that our $N$-body results can be
directly compared with the Fokker-Planck calculation of CW90.  All
parameters except for the number of stars are chosen to be the same as
used by those authors.  Models of this series have the same initial
half-mass relaxation time, and therefore all belong to CW90's family
1.  The only difference is that our models are $N$-body systems with
finite crossing time and a fully 6-dimensional phase-space
distribution of particles, while CW90 used Fokker-Planck calculations
with infinitesimal crossing time and one-dimensional distribution
function.  CW90 obtained a lifetime of 280Myr for this particular
model.  This first sequence is shown as thick solid line in Fig.\,
\ref{rgmtot}.

Models in the second series (iso-crossing, or IC, models) have the
same initial half-mass crossing time; these are similar to the runs described by
FH95.  This sequence is shown as thick dashed line in Fig.\,
\ref{rgmtot}.

\section{Results}

\subsection{Model S: the standard model}

In this section we describe the result of the standard run with 32768
particles. 

\begin{table}
\caption[]{Initial conditions and resulting dissolution time of the
main model with 32k stars.  
The first column gives the name of the
model followed by the number of stars, the tidal radius and the virial
radius (both in parsecs), the time unit in
million years and the relaxation time (in billions of years). The last
two columns contain the dissolution time in units of the time unit $\tau$
and in billions of years.'
}
\begin{flushleft}
\begin{tabular}{l|rrllr|rl} \hline
model& $N$ & $r_t$ & $r_v$ & $\tau$ & $t_{\rm rlx}$ & 
				\multicolumn{2}{c}{$t_{\rm diss}$} \\
     & &[{\rm pc}] & \unit{pc}& [Myr] & \unit{Gyr}&[$\tau$]&[Gyr]\\ \hline
S  & 32k &  62.5 & 19.94  & 7.29    & 2.87   &520 & 3.8 \\
\end{tabular}
\end{flushleft}
\label{tab:n32} \end{table}

\begin{figure}
\centerline{
\psfig{file=fig_rgmtot.ps,bbllx=570pt,bblly=40pt,bburx=110pt,bbury=690pt,height=5cm,angle=-90}}
\caption{The mass of the models plotted versus the distance to the
galactic center. The filled circles indicate the simulated models with
the life time of the model in billion years next to the symbol. 
The solid line gives the iso relaxation time models 
which are comparable to
Chernoff \& Weinberg's family 1, the dashed line gives the iso crossing
time models. 
}
\label{rgmtot}\end{figure}

\begin{figure}
\centerline{
\psfig{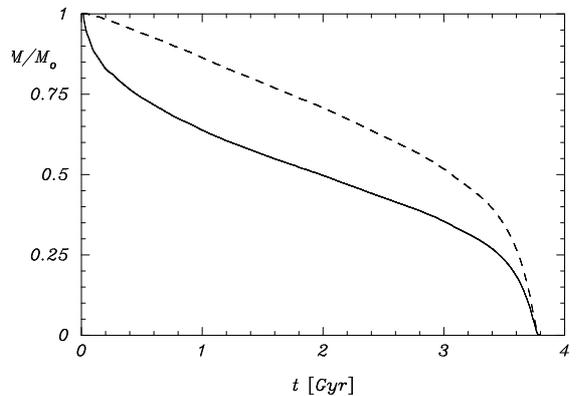}}
\caption{
The time evolution of the total mass (solid line) and number
of stars (dashed line) of the cluster for the
standard, 32768-body run. Both the mass and the number of stars are
normalized to their initial value.} 
\label{mass32krun}
\end{figure}

Figure~\ref{mass32krun} gives as a function of time (in billions of
years) the total mass and the total number of stars of the simulation
model S. This figure clearly demonstrates the initial epoch of quick
mass loss due to stellar evolution followed by a gradual decrease in
the total number of stars. Finally the stellar system evaporates.

\subsection{Models IR: constant relaxation time}

\begin{figure}
\centerline{
\psfig{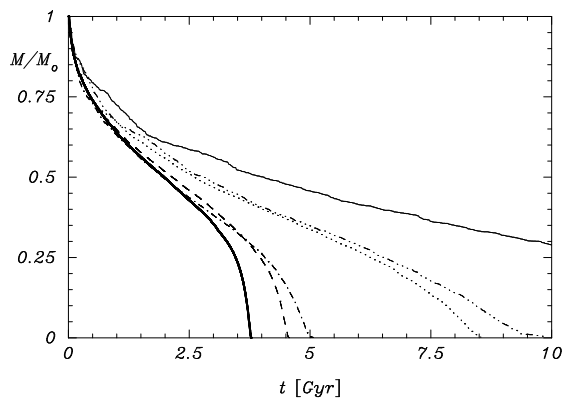}}
\caption{The mass of the star cluster as a function of time for 
the runs with constant relaxation time (models IR) of the models with 
32k (left solid line), 16k (dashed), 8k (dash-dot), 4k (dotted), 2k
(dash-3dot) and 1k (right solid line) runs.
The time scaling is cut off at 10~Gyr, at which point model IR1 has
still not dissolved completely.
}
\label{tm_irlx}\end{figure}

Table~\ref{tab:iso_relax} and Fig.\, \ref{tm_irlx} give the
results of the series of runs with constant relaxation time. 

\begin{table}
\caption[]{Initial conditions and dissolution time in the
galactic tidal field for the models with constant relaxation time.

The columns give the model name, the number of
particles, the tidal
and the virial radius (in parsec)
and the unit of time in million years. The last two columns
gives the life time of the star cluster in units of the time unit $\tau$
and in addition in billion years.
The relaxation time scale at the virial radius of these model is
approximately 2.87~Gyr.
}
\begin{flushleft}
\begin{tabular}{l|rrlr|rl} \hline
model& $N$ & $r_t$     & $r_v$     & $\tau$ & 
                 \multicolumn{2}{c}{$t_{diss}$} \\
     &     & [{\rm pc}]& \unit{pc} & [Myr]    & [$\tau$] & [Gyr]  \\ \hline
IR1  & 1k  & 151       & 47.5     & 151.8    &  220  & 33  \\ 
IR2  & 2k  & 128       & 40.5     &  84.5    &  120  &  9.9 \\ 
IR4  & 4k  & 108       & 34.8     &  47.5    &  180  &  8.5 \\
IR8  & 8k  &  90.1     & 28.1     &  24.4    &  210  &  5.1 \\ 
IR16 & 16k &  75.2     & 24.2     &  13.8    &  330  &  4.6 \\ \hline
\end{tabular}
\end{flushleft}
\label{tab:iso_relax} \end{table}

A striking feature of Fig.\, \ref{tm_irlx} is the systematic way in
which lifetimes are shorter for models with larger total mass.  This
trend can be explained as follows.  For lower mass models, holding the
relaxation time fixed implies that the crossing times are longer.
Since stars can escape from the system only on a crossing timescale at
the tidal boundary, an increase in crossing time tends to increase the 
dissolution time.  In addition, effects of stellar evolution make
themselves felt only on a crossing time, which again tends to lengthen 
the dissolution time for lower mass models.

The latter effect is most notable during the first $\sim 10^8$ years,
when the mass of the stellar system decreases dramatically. This
initial mass decrease is the result of the presence of massive stars
which evolve quickly. After the system has lost about 20\% of its
initial mass escaping stars become the major channel through which
mass is lost until the system dissolves.

By the time half of the mass has been lost, these above arguments,
related to the length of the crossing time, become less important,
since the time scale for significant change to take place has become
longer.  The subsequent difference in behavior is linked to the onset of
an instability, noted already by FH95, that operates only for high
number of stars, where an increase of the ratio $r_{\rm vir}/r_t$
leads to a loss of virial equilibrium, and a consequent rapid
dissolution of the cluster.

The strong dependence of the dissolution time on the crossing time
implies that we cannot easily extrapolate the result obtained from
small-$N$ runs to the evolution of globular clusters with realistic
number of particles. From Fig.\, \ref{tm_irlx}, it is not clear
whether or not the lifetime is converging to a particular value.  One
would hope that, in the limit of $N\rightarrow \infty$, the result
would converge to the result of the Fokker-Planck calculation.  Even
so, we have to face the question whether real globular clusters
contain a large enough number of stars to be modeled by Fokker-Planck
calculations.  So far, we have used up to 32,768 particles, a 
larger number than any previous $N$-body simulations intended to model
globular clusters.  The separation between the relaxation timescale
and the crossing timescale is quite large (more than two orders of
magnitude), and the lifetime of the system is measured in hundreds of
the crossing times. Even so, we still see a fairly strong dependence on
the crossing time for the evolution timescale of the total cluster.

In the next subsection, we examine the other way to adjust the
timescales, in the series of models with constant crossing times. 

\subsection{Models IC: constant crossing time}

\begin{figure}
\centerline{
\psfig{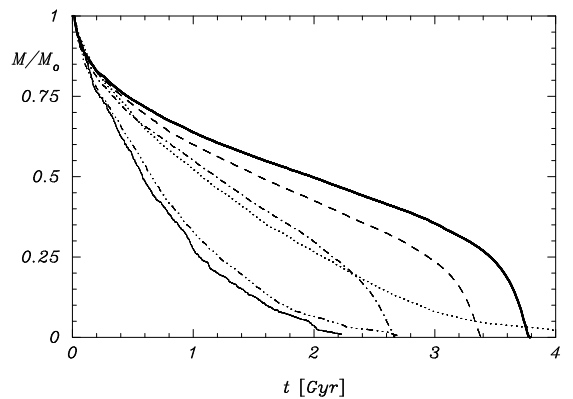}}
\caption{Mass (normalized to unity) as a function of time (in
billion years) for the computations of model IC1 (lower solid line) to
IC32 (upper solid line). Line styles are the same as in
Fig.~\ref{tm_irlx}.  The lines for the models IC5 to IC7 are not
presented in this figure.
Note the different scale along the time axis compared to
Fig.~\ref{tm_irlx}. 
}
\label{tm_idyn}\end{figure}

Table \ref{tab:iso_cross} and Fig.\, \ref{tm_idyn} shows the result of
the runs with constant crossing time. Here, the models with $N \le
4096$ and those with $N\ge 8192$ behave differently, in the sense that
the latter models show quick disruption at the end, while the former
models do not. Within the models which show the same qualitative
behavior, models with longer relaxation time evolve more slowly.

It is natural that systems with shorter relaxation times evolve
faster. If the effect of the stellar evolution is not dominant, the
main mechanism which drives the evolution of the cluster is two-body
relaxation. Thus, the evolution timescale is determined by the
relaxation timescale.

\begin{table}
\caption[]{Initial conditions and resulting lifetimes of model clusters
where the stellar evolution time is
scaled to the 
unit of time of model S ($\tau = 7.31$Myr).
The first two columns gives the model name and number of stars.
The third and fourth columns give the initial tidal radius and 
virial radius in parsec followed by the relaxation time (in
million years).
The last two columns give the dissipation time scale in units of 
$\tau$ and in billion years.
}
\begin{flushleft}
\begin{tabular}{l|rrrr|rl} \hline
model& $N$ & $r_t$ &$r_v$     & $t_{\rm rlx}$ &\multicolumn{2}{c}{$t_{diss}$} \\
     &     & \unit{pc}&\unit{pc}& \unit{Myr} & [$\tau$] & \unit{Gyr}
							\\ \hline
IC1  & 1k  & 19.69 & 6.51     &142  & 290 & 2.2 \\ 
IC2  & 2k  & 24.80 & 8.02     &250  & 370 & 2.7 \\
IC4  & 4k  & 31.28 & 9.99     &450  & 570 & 4.2 \\
IC5  & 5k  & 33.18 &10.7      &544  & 570 & 4.2 \\ 
IC6  & 6k  & 35.26 &11.4      &641  & 650 & 4.7 \\
IC7  & 7k  & 37.23 &12.1      &740  & 430 & 3.2 \\
IC8  & 8k  & 39.40 &12.7      &843  & 360 & 2.7 \\ 
IC16 & 16k & 49.63 &15.6      &1510 & 460 & 3.4 \\ \hline 
\end{tabular}
\end{flushleft}
\label{tab:iso_cross} \end{table}

\section{Discussion}

Chernoff \& Weinberg obtained a lifetime of 280 Myr for stellar systems
with a relaxation time as is chosen in the models S and IR1 to IR16.
None of our models disrupts within such a short time span, but the
trend of shorter life time for larger mass cluster is clearly visible.
However, it seems somewhat unlikely that real globular clusters could
have a disruption time of a few hundred Myrs, if we extrapolate our
numerical result. The largest number of particles used in our
calculations already reaches within a factor of 5 of that of the
smaller globular clusters.

As we stated earlier, this rather large discrepancy between the result
of our $N$-body calculation and the result of the Fokker-Planck
calculation is surprising. There are several reasons which would cause
the evaporation of the Fokker-Planck model to be different from that
of $N$-body system.  For example, the Fokker-Planck calculation relies
on the assumption of the adiabatic response of the orbits of stars to
a change of mass of the stars, assumed to be slow. This assumption is
violated for the early stage of evolution, where the stellar
evolution time scales are as short as a few Myr.

Another difference is that, in Fokker-Planck calculations, a simple
tidal limit value in energy space is used; stars with energy exceeding
this value disappear from the cluster. In our study, we remove stars
when they reach the tidal radius.  Apart from the fact that these two
procedures are already different, neither of them are appropriate. In the
Fokker-Planck calculation, both the anisotropy and the non-spherical
nature of the system, both which might have the effect of
significantly enhancing the stellar escape rate, are ignored. Our
simple treatment of the tidal boundary allows direct comparison with
CW90, but it also has the effect of reducing the escape rate, as no
external tidal force is applied to individual stars.  This is the main
reason why our test models obtained significantly longer life times than
those obtained by FH95.

In order to investigate the reason for the discrepancy between
different models, both the $N$-body calculation and Fokker-Planck
calculation have to be refined. On the side of the Fokker-Planck
calculation, until recently, further refinement has been difficult,
since except for Monte-Carlo models no practical
implementation for anisotropic models was available. However, recent
progress in the two-dimensional Fokker-Planck calculation code
(Takahashi 1993, 1995,
1996)\nocite{1996PASJ...48..691T}\nocite{1995PASJ...47..561T}\nocite{1993PASJ...45..789T}
has made the study of the effects of anisotropy feasible.

On the $N$-body side, it is fairly straightforward to try various
models for the tidal field, from the simplest one used in our present
study to the realistic static model used in FH95 (see
Fig.\,\ref{FH95_8k_tide}). 
It is even possible to go further to include the dynamic
effects of the galactic disk and the bulge. Thus, a more detailed
comparison may be possible (see the fascinating ``collaborative
experiment'' reported by Heggie, in preparation).

However, even if the $N$-body and Fokker-Planck calculations treat the
tidal boundary and the anisotropy in the same way, the difference in
the dynamical timescale still remains.  For the next several years, we
will not yet be able to model real globular clusters accurately, since
they will continue to fall in between the Fokker-Planck calculations
(with an infinitesimal crossing time) and $N$-body calculations (with
too long a crossing time). We thus urgently need some way to
interpolate between the two types of results.

The main purpose of the present study is to investigate whether such
an extrapolation is feasible. We must conclude that, as yet, there is
no obvious way to extrapolate the results from small $N$ values
upward. The models with constant relaxation time evolve too
slowly, because the crossing time is un-physically long. The models
with constant crossing time, on the other hand, evolve too
rapidly, because the relaxation timescale is too short. Although we
knew these two effects to be qualitatively present, it came to us as a 
bit of a surprise to see just how important they are, quantitatively.

Consequently, it would be practically impossible to predict the result
of a 32k run, from either of the 4k runs, the one from the IC series,
as well as the one from the IR series.  Even having both in hand would
make an extrapolation still dubious. 
This suggests that for the selected set if initial conditions an
extrapolation to real globular clusters, with hundreds of thousands of
stars, is still out of the question.

\section{Conclusions}

We have followed the evolution of a star cluster, to the point of
dissolution in the tidal field of the parent galaxy, taking into
account both the effects of stellar dynamics and of stellar evolution.
Our calculations are based on direct $N$-body integration, coupled to
approximate treatments of stellar evolution.

Our results differ greatly from those obtained with Fokker-Planck
calculations, as presented by CW90: their model clusters dissolve after
a few times $10^8$ years, whereas our equivalent model clusters live
at least ten times longer.  As we discussed in the previous section, a 
number of different reason conspire to produce such a drastic difference.

Our hope was that we would be able to find a way to bridge our
$N$-body results and previous results based on Fokker-Planck
approximations.  The fact that the GRAPE-4 special-purpose hardware
allowed us to model much larger numbers of particles, reaching to
within an order of magnitude of that of real globular clusters, seemed
to indicate that it would finally be possible to make a firm
connection between the two types of simulations.  However, our results 
indicate that no clear process of extrapolation has emerged yet.
Even within the different runs we have studied, extrapolation from the 
smaller to the larger number of stars would have resulted in rather
large errors.  This suggests that further extrapolation will suffer
from the same fate.

In the present paper, we have studied in detail a single model.
However, the way the result depends on the time scaling might be
different for other models.  In a subsequent paper, we plan to carry
out a systematic study, similar to the one we have presented here, for
a much wider range of initial models.

\acknowledgements We would like to thank Peter Eggleton, Douglas
Heggie and Koji Takahashi for many stimulating discussions.  This work
was supported in part by the Netherlands Organization for Scientific
Research (NWO) under grant PGS 78-277 and SIR~13-4109 and by the Leids
Kerkhoven Boscha Fonds.  SPZ thanks the Institute for Advanced Study
and the University of Tokyo for their hospitality.  Edward P.J.\ van
den Heuvel of the Astronomical Institute ``Anton Pannekoek'' is
acknowledged for financial support and for inviting our group for an
extended work visit.  This investigation is supported by Spinoza grant
08-0 to E.~P.~J.~van den Heuvel.


\end{document}